\documentclass[preprint,aps,prd,nofootinbib,preprintnumbers,amsmath,amssymb]{revtex4}
\usepackage{slashed}

\newcommand\fft[2]{\frac{#1}{#2}}
\newcommand\ft[2]{\tfrac{#1}{#2}}
\newcommand{\btop}[2]{\genfrac{[}{]}{0pt}{}{\,#1\,}{\,#2\,}}
\newcommand\nn{{\nonumber}}

\newcommand{\beq}{\begin{equation}}
\newcommand{\eq}{\end{equation}}
\newcommand{\bea}{\begin{eqnarray}\displaystyle}
\newcommand{\ea}{\end{eqnarray}}

\newcommand{\A}{\mathbb A}
\newcommand{\hp}{\hat\psi}

\newcommand{\tl}{\tilde{\lambda}}
\newcommand{\ma}{m=11/2}
\newcommand{\mb}{m=-9/2}

\begin{document}
\preprint{MCTP-10-45}

\title{Supersymmetric massive truncations of IIB supergravity on
Sasaki-Einstein manifolds}

\author{James T.~Liu}
\email{jimliu@umich.edu}
\affiliation{Michigan Center for Theoretical Physics,
Randall Laboratory of Physics,
The University of Michigan,
Ann Arbor, MI 48109--1040, USA}

\author{Phillip Szepietowski}
\email{pszepiet@umich.edu}
\affiliation{Michigan Center for Theoretical Physics,
Randall Laboratory of Physics,
The University of Michigan,
Ann Arbor, MI 48109--1040, USA}

\author{Zhichen Zhao}
\email{zhichen@umich.edu}
\affiliation{Michigan Center for Theoretical Physics,
Randall Laboratory of Physics,
The University of Michigan,
Ann Arbor, MI 48109--1040, USA}

\begin{abstract}
Motivated by recent interest in applications of the AdS/CFT correspondence to
condensed matter applications involving fermions, we present the
supersymmetric completion of the recent massive truncations of IIB
supergravity on Sasaki-Einstein manifolds. In particular, we
reduce the fermionic sector of IIB supergravity to obtain five
dimensional $\mathcal N =2$ supergravity coupled to one hypermultiplet and one
massive vector multiplet. The supersymmetry transformations and
equations of motion are presented and analyzed. Finally, a
particularly interesting truncation to $\mathcal N=2$ supergravity
coupled to a single hypermultiplet is presented which is the
supersymmetric completion of the recently constructed bosonic theory
dual to a 3+1 dimensional system exhibiting a superconducting phase
transition.
\end{abstract}

\maketitle

\section{Introduction}

While the study of consistent Kaluza-Klein truncations has a rich
history, until now much of it has focused on reductions that retain only
the massless sector of the lower dimensional theory.  This may be partially
attributed to the standard lore that it would be inconsistent to retain a
finite number of states in the Kaluza-Klein tower without pulling in the
rest.  However, a simple method of evading this difficulty is to retain
only singlets of a transitively acting subgroup of the internal symmetry
group.  A simple example of this prescription is the inclusion of breathing
and possibly squashing modes, and for sphere compactifications the reductions
were explicitly constructed in \cite{Bremer:1998zp}.

The basic breathing mode compactification of \cite{Bremer:1998zp} was
obtained by truncating to singlets on spheres.  Hence the resulting theories
were necessarily non-supersymmetric.  Nevertheless, as long as the underlying
theories (such as IIB or $D=11$ supergravity) are supersymmetric,
it would still be fair to investigate the supersymmetry of breathing mode
backgrounds.  This was carried out in \cite{Liu:2000gk}, where it was
demonstrated that the original fermionic variations of IIB and $D=11$
supergravity reduce to their effective $\mathcal N=2$ counterparts in five
dimensions (for IIB supergravity on a squashed $S^5$) and four dimensions
(for $D=11$ supergravity on a squashed $S^7$), respectively.  In fact,
in both cases it was possible to read off an effective $\mathcal N=2$
superpotential from the lower dimensional gravitino variations.  Moreover
it was conjectured in \cite{Liu:2000gk} that supersymmetric consistent
truncations may be obtained by retaining singlets not under the full
isometry groups SO(8) and SO(6), but rather subgroups SU(4) and SU(3),
for the squashed $S^7$ and $S^5$, respectively.

The consistent truncation conjecture of \cite{Liu:2000gk} was subsequently
verified for $D=11$ supergravity on a squashed Sasaki-Einstein manifold
by explicit construction in \cite{Gauntlett:2009zw}.  The reduction was
performed by writing $SE_7$ as U(1) bundled over a Kahler-Einstein base
and then expanding the four-form field strength in a basis of invariant
tensors corresponding to the SU(3) structure on the base.  The closure of
the SU(3) structure equations then ensures the consistency of the
truncation.  More recently, similar constructions have been obtained
for IIB supergravity on $SE_5$
\cite{Cassani:2010uw,Liu:2010sa,Gauntlett:2010vu,Skenderis:2010vz}
and $T^{1,1}$ \cite{Cassani:2010na,Bena:2010pr}.  A curious feature of these
reductions is that, while the standard reduction of IIB on $SE_5$ yields
ordinary $\mathcal N=2$ gauged supergravity in five dimensions, the massive
truncation retains a massive gravitino multiplet, and hence ought to be
viewed as a spontaneously broken $\mathcal N=4$ theory.

While the consistent truncation procedure in these cases is guaranteed to
preserve supersymmetry, until now much of the focus has been on the bosonic
sectors.  Nevertheless it would be useful to have an explicit realization
of the fermion reduction as well.  This is especially interesting in light of
holographic models of superconductivity in 2+1
\cite{Gauntlett:2009dn,Gauntlett:2009bh} and 3+1 \cite{Gubser:2009qm}
dimensions, where electronic properties often involve fermion correlators
and not just the bosons.  Along these lines, the fermion sector of the
reduction of $D=11$ supergravity on squashed $SE_7$ was recently constructed
in \cite{Bah:2010yt}.  The procedure is similar to that used in the bosonic
reduction.  In particular, the eleven-dimensional fermions may be expanded
in terms of invariant tensors multiplying Killing spinors.  This naturally
retains the lowest modes in spinor harmonics in each of the Kaluza-Klein
towers, and ensures the overall consistency of the reduction.

In this paper, we focus on the $\mathcal N=2$ truncation of IIB supergravity
reduced on squashed $SE_5$, and demonstrate the consistent reduction of
the fermion sector, at least to quadratic order in the fermions.  As
demonstrated in 
\cite{Cassani:2010uw,Liu:2010sa,Gauntlett:2010vu,Skenderis:2010vz},
the full bosonic sector of this reduction corresponds to an $\mathcal N=4$
theory.  However, by truncating out the $\mathcal N=2$ massive gravitino
multiplet, we may bring this down to $\mathcal N=2$.  While our main
motivation for doing so is to avoid unnecessarily cumbersome expressions
related to the massive gravitino sector, we do not see any obstacles to
achieving the full reduction if desired.  Furthermore, this allows us to
highlight some of the features of the reduction from an $\mathcal N=2$
perspective.

Since the reduction of the fermionic sector uses the bosonic reduction as a
starting point, we begin with a brief review of the bosons in
Section~\ref{sec:IIBred}.  We then turn to the reduction of the IIB fermions
in Section~\ref{sec:fermion} and present the effective five-dimensional
theory in Section~\ref{sec:five}.  Moreover, as shown in Section~\ref{sec:sc},
the resulting $\mathcal N=2$ theory admits a truncation to gauged supergravity
coupled to a single hypermultiplet, corresponding to the model of
\cite{Gubser:2009qm} for a holographic superconductor in 3+1 dimensions.

While this work was being completed we became aware of \cite{Bah:2010cu},
which has substantial overlap with our results.  In fact, \cite{Bah:2010cu}
has worked out the reduction of the fermions corresponding to the
full $\mathcal N=4$ theory, thus demonstrating consistency of the complete
massive truncation, and not just the $\mathcal N=2$ sector that we focus on
here.

\section{The bosonic reduction of IIB supergravity on $SE_5$}
\label{sec:IIBred}

The reduction of the bosonic sector of IIB supergravity on a squashed
Sasaki-Einstein manifold was carried out in 
\cite{Cassani:2010uw,Liu:2010sa,Gauntlett:2010vu,Skenderis:2010vz}.  From
an $\mathcal N=2$ point of view, the resulting theory has on-shell fields
corresponding to that of five-dimensional gauged supergravity coupled to a
massive hypermultiplet, massive gravitino multiplet and massive vector
multiplet \cite{Liu:2000gk,Liu:2010sa}.

Before turning to the fermions, we review the reduction of the bosonic
sector, following the notations and conventions of \cite{Liu:2010sa}.
Although IIB supergravity does not admit a covariant action, we may
take a bosonic Lagrangian of the form
\begin{equation}
\mathcal L_{\mathrm{IIB}}=R*1-\fft1{2\tau_2^2}d\tau\wedge*d\bar\tau
-\fft12\mathcal M_{ij}F_3^i
\wedge*F_3^j-\fft14\widetilde F_5\wedge*\widetilde F_5-\fft14\epsilon_{ij}
C_4\wedge F_3^i\wedge F_3^j,
\label{eq:iiblag}
\end{equation}
where self-duality $\widetilde F_5=*\widetilde F_5$ is to be imposed by
hand after deriving the equations of motion.  Here we have chosen to write
the Lagrangian in an SL(2,$\mathbb R$) invariant form using
\begin{equation}
\tau=C_0+ie^{-\phi},\qquad
\mathcal V=\fft1{\sqrt{\tau_2}}\begin{pmatrix}-\tau_1&1\cr\tau_2&0
\end{pmatrix},\qquad
\mathcal M=\mathcal V^T\mathcal V=
\fft1{\tau_2}\begin{pmatrix}|\tau|^2&-\tau_1\cr-\tau_1&1
\end{pmatrix}.
\end{equation}
For convenience when coupling to fermions, we also introduce the complexified
vielbein $v_i=\mathcal V^1{}_i-i\mathcal V^2{}_i$, so that
\begin{equation}
v_iF_3^i=\tau_2^{-1/2}(F_3^2-\tau F_3^1)=\tau_2^{-1/2}G_3,
\label{eq:complexv}
\end{equation}
where $G_3=F^2_3-\tau F^1_3$.

The reduction ansatz follows by taking a metric of the squashed Sasaki-Einstein
form
\begin{equation}
ds_{10}^2=e^{2A}ds_5^2+e^{2B}ds^2(B)+e^{2C}(\eta+A_1)^2,
\label{eq:metans}
\end{equation}
where $d\eta=2J$ and where we set $3A+4B+C=0$ to remain in the Einstein
frame.  The key to the
reduction is to expand the remaining bosonic fields in terms of the
invariant forms $J$ and $\Omega$ based on the SU(2) structure of the base
$B$ and satisfying
\begin{equation}
J\wedge\Omega=0,\qquad \Omega\wedge\bar\Omega=2J\wedge J=4*_41,\qquad
*_4J=J,\qquad*_4\Omega=\Omega,
\end{equation}
as well as
\begin{equation}
dJ=0,\qquad d\Omega=3i(d\psi+\mathcal A)\wedge\Omega.
\end{equation}

The bosonic reduction follows by expanding the three-form and five-form
field strengths in a basis of invariant tensors on $B$.  Since we will
truncate out the massive gravitino multiplet, we set the corresponding
bosonic fields to zero. (The complete reduction is given in
\cite{Liu:2010sa}.)  In this case, the three-form gives rise to two
complex scalars $b^i$, and is given by
\begin{equation}
F_3^i=f_1^i\wedge\Omega
+\bar f_1^i\wedge\bar\Omega+f_0^i\wedge\Omega\wedge(\eta+A_1)
+\bar f_0^i\wedge\bar\Omega\wedge(\eta+A_1),
\label{eq:f3ans}
\end{equation}
where
\begin{equation}
f_1^i=Db^i,\qquad f_0^i=3ib^i,
\end{equation}
with $D$ the U(1) gauge covariant derivative
\begin{equation}
Db^i=db^i-3iA_1b^i.
\end{equation}
Furthermore, introducing
\begin{equation}
b^i=\begin{pmatrix}1\cr\tau\end{pmatrix}b^{m^2=-3}+
\begin{pmatrix}1\cr\bar\tau\end{pmatrix}b^{m^2=21},
\end{equation}
it is easy to see that
\begin{equation}
v_if_0^i=6\sqrt{\tau_2}\,b^{m^2=21},\qquad
\bar v_if_0^i=-6\sqrt{\tau_2}\,b^{m^2=-3},
\label{eq:vfexp}
\end{equation}
while
\begin{eqnarray}
v_if_1^i&=&-2i\sqrt{\tau_2}[Db^{m^2=21}+\ft{i}{2\tau_2}(b^{m^2=-3}d\tau
+b^{m^2=21}d\bar\tau)],\nonumber\\
\bar v_if_1^i&=&2i\sqrt{\tau_2}[Db^{m^2=-3}-\ft{i}{2\tau_2}(b^{m^2=-3}d\tau
+b^{m^2=21}d\bar\tau)].
\end{eqnarray}
These expressions will show up extensively in the fermion reduction below.

For the self-dual five-form, we have
\begin{equation}
\widetilde F_5=(1+*)[(4+\phi_0)*_41\wedge(\eta+A_1)
+\mathbb A_1\wedge*_41+p_2\wedge J\wedge(\eta+A_1)],
\end{equation}
where $*_41$ denotes the volume form on the Kahler-Einstein base $B$.
The fields $\phi_0$ and $p_2$ are constrained by
\begin{eqnarray}
\phi_0&=&-\ft{2i}3\epsilon_{ij}(f_0^i\bar f^j_0-\bar f^i_0f^j_0),
\nonumber\\
p_2&=&-d[A_1+\ft14\mathbb A_1+\ft{i}6\epsilon_{ij}(f_0^i\bar f^j_1
-\bar f^i_0f^j_1)].
\label{eq:constraints}
\end{eqnarray}
Hence the only additional field arising from the five-form is the vector
$\mathbb A_1$.

Finally, we note that the bosonic field content of this massive truncation
is that of gauged supergravity coupled to a hypermultiplet with fields
$(\tau,b^{m^2=-3})$ and a massive vector multiplet with fields
$(B,C,b^{m^2=21},\mathbb A_1)$.  This massive multiplet is actually a vector
combined with a hypermultiplet.  However, since we are working on shell,
one of the scalars has been absorbed into the massive vector.
If desired, this scalar may be restored by an appropriate Stueckelberg shift
of $\mathbb A_1$.

\section{Reduction of the IIB fermions}
\label{sec:fermion}

We are now prepared to examine the fermionic sector of IIB supergravity
\cite{Schwarz:1983qr}.  For simplicity in working out the reduction, we
follow a Dirac convention throughout.  In this case, the fermions consist of
a spin-$\fft32$ gravitino $\Psi_M$ and a spin-$\fft12$ dilatino $\lambda$,
with opposite chiralities
\beq
\Gamma_{11} \Psi_M = \Psi_M, \qquad \Gamma_{11}\lambda = -\lambda.
\eq
Our Dirac conventions are detailed in Appendix \ref{app:dirac}.  In
particular, as opposed to \cite{Schwarz:1983qr}, we are using a mostly plus
metric signature.

In the following we always work to lowest order in the fermions.  In
this case, the IIB supersymmetry variations on the fermions are given by
\cite{Schwarz:1983qr}
\bea \label{eq:susyvars}
\delta\lambda &=& \frac{i}{2\tau_2}\Gamma^A\partial_A\tau\epsilon^c
-\frac{i}{24}\Gamma^{ABC}v_iF^i_{ABC}\epsilon, \nn \\ 
\delta\Psi_M &=&\mathcal D_M\epsilon\equiv
\left(\nabla_M + \frac{i}{4\tau_2}\partial_M\tau_1 +
\frac{i}{16\cdot5!} \Gamma^{ABCDE}\widetilde F_{ABCDE}\Gamma_M\right)
\epsilon \nn\\ 
&&\kern4em +\frac{i}{96}\left(\Gamma_M{}^{ABC}-9\delta_M^A\Gamma^{BC}\right)
v_iF^i_{ABC}\epsilon^c.
\ea
The supersymmetry parameter
$\epsilon$ is chiral with $\Gamma_{11}\epsilon =\epsilon$, and the
complexified $SL(2,\mathbb R)$ vielbein, $v_i$, was defined above in
(\ref{eq:complexv}).  In addition the fermion equations of motion are
\cite{Schwarz:1983qr}
\begin{eqnarray}
0&=&\Gamma^M\mathcal D_M\lambda-\fft{i}{8\cdot5!}\Gamma^{MNPQR}F_{MNPQR}
\lambda, \nn \\
0&=&\Gamma^{MNP}\mathcal D_N\Psi_P
+ \fft{i}{48}\Gamma^{NPQ}\Gamma^M
v_i^*F^{i*}_{NPQ}\lambda - \fft{i}{4\tau_2}\Gamma^N\Gamma^M\partial_N
\tau\lambda^c,
\label{eq:iibeom}
\end{eqnarray}
where the supercovariant derivative acting on the gravitino is defined in
the gravitino variation (\ref{eq:susyvars}).  On the other hand, the
supercovariant derivative acting on the dilatino takes the form
\begin{equation}
\mathcal D_M\lambda=\left(\nabla_M + \fft{3i}{4\tau_2}\partial_M\tau_1\right)
\lambda
- \fft{i}{2\tau_2}\Gamma^N\partial_N\tau\Psi_M^c
+ \fft{i}{24}\Gamma^{NPQ}v_iF^i_{NPQ}\Psi_M,
\end{equation}
and is defined so that $\nabla_M\epsilon$ terms drop out of the variation
$\mathcal D_M\delta\lambda$, as appropriate to supercovariantization.

\subsection{Killing spinors on $SE_5$}

The starting point of the fermion reduction is the construction of Killing
spinors on $SE_5$.  Starting with the undeformed Sasaki-Einstein metric
\begin{equation}
ds^2(SE_5)=ds^2(B)+(d\psi+\mathcal A)^2,
\end{equation}
the Killing spinor equations then follow from the internal components
of the gravitino variation in (\ref{eq:susyvars}) with a constant
five-form flux
\begin{equation}
\tilde F_5 = 4*_51 + 4*_41\wedge (d\psi +\mathcal A)
\end{equation}
and take the form
\begin{eqnarray}
0&=&\delta\Psi_a=\hat{\mathcal D}_a\eta\equiv
[\hat\nabla_a-\mathcal A_a\partial_\psi+\ft12J_{ab}\tau^b
\tau^9+\ft{i}2\tau_a]\eta,\nonumber\\
0&=&\delta\Psi_9=[\partial_\psi-\ft14J_{ab}\tau^{ab}+\ft{i}2\tau_9]\eta.
\label{eq:sekse}
\end{eqnarray}
We proceed by assigning a U(1) charge $q$ to the Killing spinor $\eta$, so
that $\partial_\psi\eta=iq\eta$.  Furthermore, since $(J_{ab}\tau^{ab})^2
=-8(1-\tau^9)$, we see that $J_{ab}\tau^{ab}$ has eigenvalues $(4i,-4i,0,0)$
with corresponding $\tau^9$ eigenvalues $(-1,-1,1,1)$.  The variation
$\delta\Psi_9$ then vanishes for the charges
$q=(\fft32,-\fft12,-\fft12,-\fft12)$.  The $\mathcal N=2$ Killing spinor
is thus obtained by taking $q=\fft32$ and $J_{ab}\tau^{ab}\eta=4i\eta$.

Having exhausted the content of the $\delta\Psi_9$ equation, we now turn
to integrability of $\delta\Psi_a$, which gives the requirement
\begin{equation}
0=\tau^b[\hat{\mathcal D}_a,\hat{\mathcal D}_b]\eta=\tau^b
[\delta_{ab}(\tau^9-1)-iJ_{ab}(\tau^9+2q)]\eta.
\end{equation}
For $q=\fft32$ and $\tau^9\eta=-\eta$, this gives the condition
$J_{ab}\tau^b\eta=i\tau_a\eta$, which is easily seen to be consistent
with the above requirement that $J_{ab}\tau^{ab}\eta=4i\eta$.  After
defining $\eta = e^{3i\psi/2}\tilde\eta$, we are finally left with the
condition
\begin{equation}
[\hat\nabla_a-\ft{3i}2\mathcal A_a]\tilde\eta=0,
\end{equation}
which is solved by taking $\tilde\eta$ to be a gauge covariantly constant
spinor on the Kahler-Einstein base \cite{Gibbons:2002th}.

To summarize the above, the system (\ref{eq:sekse}) may be solved to yield
a single complex Killing spinor $\eta$ satisfying
\begin{equation}
\label{eq:ksproj}
\partial_\psi\eta=\ft{3i}2\eta,\qquad
\tau^9\eta = -\eta, \qquad  \tau^bJ_{ab} \eta = i \tau_a \eta, \qquad
\tau^{b}\Omega_{ab} \eta = 0.
\end{equation}
The final condition may be obtained by multiplying the penultimate one
by $\Omega_{ca}$ on both sides and making use of the identity
$\Omega_{ca}J_{ab}=-i\Omega_{cb}$, which follows from the relations
\cite{Gauntlett:2010vu}
\begin{equation}
\Omega_{ac}\Omega^{bc}=0, \qquad
\Omega_{ac}\bar\Omega^{bc}=2\delta_a{}^b - 2iJ_a{}^b.
\end{equation}
The Killing spinor $\eta$ and its conjugate $\eta^c$ provide a natural
basis of invariant spinors in which to expand the fermions.  Furthermore,
as discussed in \cite{Bah:2010yt}, these represent singlets of the $SU(2)$
structure group, thus ensuring consistency of the reduction.  Note that
$\eta$ and $\eta^c$ are related by
\begin{equation}
\tau^{b}\bar\Omega_{ab}\eta =2\tau_a \eta^c,
\end{equation}
and $\eta^c$ satisfies the conjugated relations
\begin{equation}
\partial_\psi\eta^c=-\ft{3i}2\eta^c,\qquad
\tau^9\eta^c = -\eta^c, \qquad  \tau^bJ_{ab} \eta^c = -i \tau_a \eta^c, \qquad
\tau^{b}\bar\Omega_{ab} \eta^c = 0.
\end{equation}

\subsection{IIB spinor decomposition} \label{sec:spinordecomp}

We are now in a position to present the fermion decomposition ansatz by
expanding the ten-dimensional fermions in terms of $\eta$ and $\eta^c$.
Although we will ultimately truncate away the massive gravitino multiplet,
we find it instructive to start with the complete ansatz.  This allows us
to identify which fermions belong in which multiplets, and hence will guide
the truncation.

Starting with the IIB dilatino, since it has negative chirality, it may
be decomposed as%
\footnote{Note that this is a slight abuse of
notation, in that $\lambda$ shows up as both ten-dimensional and
five-dimensional fields.  The correct interpretation will be obvious from
the context.}
\beq
\lambda = e^{-A/2}\lambda\otimes\eta\otimes\btop01 +
e^{-A/2}\lambda'\otimes\eta^c\otimes\btop01.
\label{eq:lamans}
\eq
The IIB transformation parameter $\epsilon$ and gravitino $\Psi_A$ each have
positive chirality.  Thus we expand the gravitino in ten dimensional flat
indices as
\bea
\Psi_\alpha &=& e^{-A/2}\psi_\alpha\otimes\eta\otimes \btop10
 + e^{-A/2}\psi'_\alpha\otimes\eta^c\otimes\btop10, \nn \\
\Psi_a &=& e^{-A/2}\psi \otimes\tau_a\eta\otimes\btop10
 + e^{-A/2}\psi'\otimes\tau_a\eta^c\otimes\btop10, \nn \\
\Psi_9 &=& e^{-A/2}\psi_9\otimes\tau_9\eta\otimes\btop10
 + e^{-A/2}\psi'_9\otimes\tau_9\eta^c\otimes\btop10,
\label{eq:psians}
\ea
and the transformation parameter as
\beq
\epsilon = e^{A/2}\varepsilon\otimes\eta\otimes\btop10.
\label{eq:epsans}
\eq
Note that in all the above we have included relevant warp factors to
account for the breathing and squashing modes.

While we have started with a theory with 32 real supercharges, only a quarter
of these are preserved in the AdS$_5\times SE_5$ background.  By focusing
on supersymmetries generated by (\ref{eq:epsans}), we are thus restricting
our study to five-dimensional supersymmetry parameterized by a single Dirac
spinor.  This corresponds to an $\mathcal N=2$ theory, and provides a
motivation for us to remove the massive gravitino from subsequent consideration.
(If desired, the full spontaneously broken $\mathcal N=4$ symmetry may be
obtained by introducing an $\varepsilon\otimes\eta^c$ component in
(\ref{eq:epsans}).  However, we will not pursue this here.)

\subsection{Linearized analysis and the $\mathcal{N} = 2$ supermultiplet
structure}\label{sec:lin}

Before presenting the fermionic reduction, it is instructive to analyze
the linearized equations of motion. Doing so allows us to group the effective
five-dimensional fermions into the relevant $\mathcal N=2$
supermultiplets as highlighted in \cite{Liu:2010sa}.  We start by noting
that the five-dimensional fermions consist of the two gravitini $\psi_\alpha$
and $\psi_\alpha'$, two dilatini $\lambda$ and $\lambda'$ and four additional
spin-1/2 fields $\psi$, $\psi'$, $\psi_9$ and $\psi_9'$ arising from the
internal components of the ten-dimensional gravitino.

In the linearized theory, the equations are greatly simplified and the
fermions satisfy free massive Dirac and Rarita-Schwinger equations.  
The $\lambda$ and $\lambda'$ equations are naturally diagonal and the
gravitino equations are diagonalized by the following modes, 
\bea
&&\hp_\alpha = \psi_\alpha +
\ft{i}3\gamma_\alpha\left(4\psi+\psi_9\right), \kern6.5em \psi^{m=11/2} =
4\psi+\psi_9, \qquad 
 \psi^{m=-9/2} = \psi-\psi_9, \nn \\
&& \hp'_\alpha = \psi'_\alpha + \ft{i}{10}\left(\gamma_\alpha+ 2
\nabla_\alpha\right)\left(4\psi' + \psi_9'\right), \qquad \psi'^{m=5/2} =
\psi' - \psi_9'.
\label{eq:shifts}
\ea
In all, the linearized modes satisfy,
\bea
&&\gamma^{\mu\alpha\beta}\nabla_\alpha\hat\psi_\beta =
\ft32\gamma^{\mu\alpha}\hat\psi_\alpha,\kern4.5em
\gamma^{\mu\alpha\beta}\nabla_\alpha\hat\psi'_{\beta} =
-\ft72\gamma^{\mu\alpha}\hat\psi'_\alpha,\nonumber\\
&&
\gamma^\alpha\nabla_\alpha \lambda = \ft{7}2\lambda,\kern8em
\gamma^\alpha\nabla_\alpha \lambda' = -\ft{3}2 \lambda',\nonumber\\
&&
\gamma^\alpha\nabla_\alpha \psi^{m=11/2} = \ft{11}2 \psi^{m=11/2},\qquad
\gamma^\alpha\nabla_\alpha \psi^{m=-9/2} = - \ft{9}2\psi^{m=-9/2},\nonumber\\
&&\gamma^\alpha\nabla_\alpha \psi'^{m=5/2} = \ft52\psi'^{m=5/2}.
\ea
Note that the massive gravitino obtains its mass by absorbing the
spin-1/2 combination $4\psi'+\psi_9'$.

\begin{table}[t]
\begin{tabular}{l|l|l|l}
n&Multiplet&State&Field\\
\hline
0&supergraviton&$D(4,1,1)_0$&$g_{\mu\nu}$\\
&&$D(3\fft12,1,\fft12)_{-1}+D(3\fft12,\fft12,1)_1$&$\hat\psi_\mu$\\
&&$D(3,\fft12,\fft12)_0$&$A_1+\fft16\mathbb A_1$\\
\hline
0&LH+RH chiral&$D(3,0,0)_{\pm2}$&$b^{m^2=-3}$\\
&&$D(3\fft12,\fft12,0)_1+D(3\fft12,0,\fft12)_{-1}$&$\lambda'$\\
&&$D(4,0,0)_0+D(4,0,0)_0$&$\tau$\\
\hline
1&LH+RH massive gravitino&$D(5\fft12,\fft12,1)_1+D(5\fft12,1,\fft12)_{-1}$&$\hat\psi'_\mu$\\
&&$D(5,\fft12,\fft12)_0+D(5,\fft12,\fft12)_0$&$b_1^i$\\
&&$D(5,0,1)_2+D(5,1,0)_{-2}$&$q_2$\\
&&$D(6,0,1)_0+D(6,1,0)_0$&$b_2^i$\\
&&$D(4\fft12,0,\fft12)_1+D(4\fft12,\fft12,0)_{-1}$&$\psi'^{m=5/2}$\\
&&$D(5\fft12,0,\fft12)_{-1}+D(5\fft12,\fft12,0)_1$&$\lambda$\\
\hline
2&massive vector&$D(7,\fft12,\fft12)_0$&$\mathbb A_1$\\
&&$D(6\ft12,\ft12,0)_{-1}+D(6\ft12,0,\ft12)_1$&$\psi^{m=-9/2}$\\
&&$D(7\ft12,0,\ft12)_{-1}+D(7\ft12,\ft12,0)_1$&$\psi^{m=11/2}$\\
&&$D(6,0,0)_0$&$\sigma$\\
&&$D(7,0,0)_{\pm2}$&$b^{m^2=21}$\\
&&$D(8,0,0)_0$&$\rho$\\
\end{tabular}
\caption{Identification of the bosonic and fermionic states in the Kaluza-Klein spectrum
with the linearized modes in the reduction.}
\label{tbl:match}
\end{table}

As with the fields in the bosonic truncation, we have arrived at 
a field content which, in the case of the round five-sphere, saturates
the lowest harmonic in each of the respective Kaluza-Klein towers as
determined in \cite{Gunaydin:1984fk,Kim:1985ez}.  Noting that, in five
dimensions, the relation
between the conformal weight $\Delta$ and mass $m$ of the fermions is
$|m|=\Delta-2$, we can map the fermion fields into $\mathcal N=2$ AdS
multiplets. First, it is straightforward to see
that $\hat\psi_\mu$ has $m=3/2$, corresponding to a massless
spin-$3/2$ field in AdS$_5$.  Hence it should be identified with the
massless gravitino sitting in the supergraviton multiplet.
Also at the zeroth Kaluza-Klein level, the LH+RH chiral multiplet contains an
$m=3/2$ fermion which may be identified as $\lambda'$. At level $n=1$,
the massive
gravitino multiplet has three fermions; one spin-$3/2$ particle with
$m=-7/2$ corresponding to the massive gravitino $\hat\psi'_\mu$ and
two spin-$1/2$ particles with $m=5/2$ corresponding to $\psi'^{m=5/2}$
and $m=7/2$  corresponding to $\lambda$. Finally, at the $n=2$
Kaluza-Klein level, the massive vector multiplet contains two
spin-$1/2$ particles, $\psi^{m=-9/2}$ and $\psi^{m=11/2}$. These
identifications will be further justified by examining the
supersymmetry transformations. The complete field content of the
supermultiplets is shown in Table~\ref{tbl:match}, where the bosonic
fields are fully defined in \cite{Liu:2010sa}.


\section{The Five-dimensional Theory and $\mathcal N=2$ Supergravity}
\label{sec:five}

The linearized analysis above demonstrates that the fields $\psi_\alpha'$,
$\psi'$, $\psi_9'$ and $\lambda$ belong to the massive gravitino multiplet.
We thus proceed with the $\mathcal N=2$ truncation by setting these to zero
\beq
\psi'_\alpha = 0, \qquad \psi' = 0, \qquad  \psi'_9 = 0, \qquad \lambda = 0.
\eq
It is straightforward to show this this is a consistent truncation, provided
the bosonic fields in the massive graviton multiplet are set to zero%
\footnote{The consistency of this truncation in the
bosonic sector has been previously shown in
\cite{Liu:2010sa,Gauntlett:2010vu,Cassani:2010uw}.}.
Moreover, other 
than just simplifying the resulting equations, this truncation is
natural when explicitly discussing $\mathcal N=2$ supersymmetry as the
massive gravitino should really be thought of as descending from a
spontaneously broken $\mathcal N=4$ theory.

\subsection{Supersymmetry Variations}
 
We start with the reduction of the IIB supersymmetry variations given in
(\ref{eq:susyvars}). Inserting the fermion ans\"atze
(\ref{eq:lamans}), (\ref{eq:psians}) and (\ref{eq:epsans}) into the IIB
variations, we arrive at the following five-dimensional variations%
\footnote{Note that with the
Dirac matrix conventions described in the appendix we have $\epsilon^c = i
\varepsilon^c\otimes\eta^c\otimes\btop10.$}
\bea
\delta\hat\psi_\alpha&\equiv&\mathcal D_\alpha\varepsilon= \Big[D_\alpha +\ft{i}{24}e^{C-A}\left(\gamma_\alpha{}^{\nu\rho}-4\delta_\alpha{}^{\nu}\gamma^{\rho}\right) \left(F_{\nu\rho}-2e^{-2B-2C}p_{\nu\rho}\right)\nn \\
&&\kern4em + \ft1{12}\gamma_\alpha\left(4e^{A-2B+C} + 6e^{A-C} - (4+\phi_0)e^{A-4B-C}\right)\Big]\varepsilon \nn \\
\label{eq:gvar}
&&\kern4em - e^{-2B}\left(v_if_\alpha^i - \ft{i}3e^{A-C}v_if_0^i\gamma_\alpha\right)\varepsilon^c,  \\
\delta\psi^{m=11/2}\!&=&\!\Big[-\ft{i}2\gamma^\mu\partial_\mu\left(4B+C\right)- \ft38e^{-4B}\gamma^\mu\A_\mu + \ft18e^{C-A}\gamma^{\mu\nu}\left(F_{\mu\nu}+e^{-2B-2C}p_{\mu\nu}\right)-ie^{A-2B+C} \nn \\
&& -\ft{3i}2e^{A-C}+\ft{5i}8(4+\phi_0)e^{A-4B-C}\Big]\varepsilon +      
e^{-2B}\left(\ft{3i}4\gamma^\mu v_if^i_\mu + \ft{7}4e^{A-C} v_if^i_0\right)\varepsilon^c, \\
\label{eq:11/2var}
\delta\psi^{m=-9/2}\!&=&\!\Big[-\ft{i}2\gamma^\mu\partial_\mu\left(B-C\right) -\ft14e^{-4B}\gamma^\mu\A_\mu - \ft18e^{C-A}\gamma^{\mu\nu}\left(F_{\mu\nu}+e^{-2B-2C}p_{\mu\nu}\right) \nn \\
&& - \ft{3i}2e^{A-2B+C} + \ft{3i}2e^{A-C}\Big] \varepsilon  +
e^{-2B}\left(\ft{i}2 \gamma^\mu v_i f^i_\mu - \ft{1}2e^{A-C} v_i f_0^i
  \right)\varepsilon^c,  \\ 
\label{eq:9/2var}
\delta\lambda' &=& -\ft1{2\tau_2}\gamma^\mu\partial_\mu\tau\varepsilon^c -
ie^{-2B}\left(\gamma^\mu v_i \bar f_{\mu}^i - ie^{A-C}v_i\bar
f_0^i\right)\varepsilon.
\label{eq:dilvar}
\ea
The gauge covariant derivative $D_\alpha$ acting on $\varepsilon$ is given
by $D_\alpha \equiv \nabla_\alpha - \fft{3i}{2}(A_\alpha+\fft16e^{-4B}
\mathbb A_\alpha) + \fft{i}{4\tau_2}
\partial_\alpha\tau_1$, where the latter term descends from
the traditional charge with respect to the U(1) compensator field,
$Q_M$, in the ten dimensional IIB theory \cite{Schwarz:1983qr}.  Furthermore,
we have defined the five-dimensional supercovariant derivative $\mathcal
D_\alpha$ through the gravitino variation in (\ref{eq:gvar}).

There are several facts worth noting about these expressions.
Firstly, we see that these variations fit nicely into the multiplet
structure as presented in Table~\ref{tbl:match}.  In particular, the dilatino
variation is built out of $\tau$ and $\bar v_i f^i$, both of which belong
to the LH+RH chiral multiplet, since the latter corresponds to $b^{m^2=-3}$
according to (\ref{eq:vfexp}).  On the other hand, $\delta\psi^{m=11/2}$ and
$\delta\psi^{m=-9/2}$ contain only terms involving fields from the
graviton and massive vector multiplets.  [Note that the combination
$F_2+e^{-2B-2C}p_2$ appearing in (\ref{eq:11/2var}) and (\ref{eq:9/2var})
essentially selects the field strength of the massive vector
$\mathbb A_1$, as can be seen from the definition of $p_2$ given in
(\ref{eq:constraints})].  These observations give
further justification for the multiplet structure presented in section
\ref{sec:lin}.  
  
Furthermore, since the breathing mode is $\rho \sim 4B+C$, and the
squashing mode is $\sigma \sim B-C$, we can identify $\psi^{m=11/2}$
with the fermionic partner of the breathing mode and $\psi^{m=-9/2}$
as the fermionic partner of the squashing mode as first demonstrated
in \cite{Liu:2000gk}. Finally, from the gauge covariant derivative, it
is evident that the combination $A_\mu + \fft16 e^{-4B}\A_\mu$ may
be identified with the graviphoton, which is consistent with the
linearized analysis in \cite{Liu:2010sa}.  (The combination
$F_2-2e^{-2B-2C}p_2$ appearing in the gravitino variation is similarly
the effective graviphoton field strength.)

The gravitino variation (\ref{eq:gvar}) is particularly interesting, as we
may attempt to read off an $\mathcal N=2$ superpotential from the term
proportional to $\gamma_\alpha\varepsilon$
\beq
W = 2e^{A-2B+C} + 3e^{A-C}-\fft12(4+\phi_0)e^{A-4B-C}.
\eq
Recalling the relations $3A+4B+C=0$ and $\phi_0 =
-\ft{2i}3\epsilon_{ij}\left(f_0^i\bar f_0^j - \bar f_0^i
f_0^j\right)$, we see that the scalar potential can be written as
\beq \label{eq:sp}
V = 2 (\mathcal G^{-1})^{ij}\partial_i W\partial_j W - \fft43 W^2,
\eq
where $(\mathcal G^{-1})^{ij}$ is the inverse scalar metric which can
be read off from the scalar kinetic terms in the Lagrangian and
$\{i,j\}$ run over all scalars in the theory.

To verify (\ref{eq:sp}), we made use of the fact that the scalar metric
given in \cite{Liu:2010sa} is
composed of three independent components, pertaining to the
independent sets of scalars $\{B,C\}$, $\{b_0^1, b_0^2\}$ and $\tau$, with
explicit components
\beq
(\mathcal G_{\{B,C\}}^{-1})^{ij} = \fft1{16}\begin{pmatrix}1&-1\cr-1&7
\end{pmatrix}, \qquad (\mathcal G_{\{b_0^1,b_0^2\}}^{-1})^{ij} =
\fft{e^{4B}}{4\tau_2}\begin{pmatrix}1&\tau_1\cr\tau_1&|\tau|^2
\end{pmatrix}, \qquad \mathcal G_{\tau}^{-1} = \tau_2^2.  
\eq
Inserting these expressions into (\ref{eq:sp}) then exactly reproduces the
scalar potential appearing in the bosonic Lagrangian.  This is, however, a
somewhat surprising relation as the actual gravitino variation (\ref{eq:gvar})
contains not only the term proportional to the superpotential written
above, but another term involving $v_if_0^i\varepsilon^c$ where
$v_if_0^i$ is proportional to $b_0^{m^2=21}$, as indicated in (\ref{eq:vfexp}).
Based on general $\mathcal N=2$ gauged supergravity arguments, this should
conceivably also contribute to the scalar potential, but is not taken into
account by (\ref{eq:sp}).

\subsection{Equations of Motion} \label{sec:iibeom}

Turning to the equations of motion, the reduction of the dilatino equation
is the most straightforward.  After a bit of manipulation, we obtain
\bea
0 &=& \left[\gamma^\mu\mathcal D_\mu+\ft{i}8\gamma^{\mu\nu}
\left(e^{C-A}F_{\mu\nu}-2e^{-A-2B-C}p_{\mu\nu}\right)
 - \ft14(4+\phi_0)e^{A-4B-C} +
e^{A-2B+C} + \ft32e^{A-C}\right]\lambda' \nn \\
&& - e^{-2B}v_i \left[\ft45\gamma^\mu\bar f_\mu^i + \ft{28i}{15}\bar f_0^i
\right]\psi^{\ma} - e^{-2B}v_i\left[\ft45\gamma^\mu\bar f_\mu^i -
  \ft{4i}{5}\bar f_0^i e^{A-C}\right]\psi^{\mb}, 
\ea
where the supercovariant derivative acting on the dilatino is defined by
\begin{equation}
\mathcal D_\mu\lambda'\equiv D_\mu\lambda'-K(\lambda')\hat\psi_\mu
=\left[\nabla_\mu +
\ft{3i}{4\tau_2}\partial_\mu\tau_1 + \ft{3i}2\left(A_\mu +\ft16e^{-4B}
\A_\mu\right)\right]\lambda'-K(\lambda')\hat\psi_\mu.
\end{equation}
The supercovariantization term $K(\lambda')$ acting on $\hat\psi_\mu$ is
given by the right hand side of the dilatino variation (\ref{eq:dilvar})
with $\varepsilon$ replaced by $\hat\psi_\mu$ (and similarly $\varepsilon^c$
replaced by $\hat\psi_\mu^c$).

Starting with the IIB gravitino, we arrive at three equations, corresponding
to the $\alpha$, $a$, and $9$ components.  After a fair bit of manipulations,
and the appropriate redefinitions given in the first line of (\ref{eq:shifts}),
we obtain the $\psi^{m=11/2}$ and $\psi^{m=-9/2}$ equations
\bea
0 &=& \Bigl[\gamma^\mu\mathcal D_\mu + \ft{3i}{5}e^{-4B}\gamma^\mu\A_\mu
-\ft{i}{120}e^{C-A}\gamma^{\mu\nu}F_{\mu\nu} - \ft{11i}{60}e^{-A-2B-C}\gamma^{\mu\nu}p_{\mu\nu} \nn \\
&&+ e^A\left(-\ft{17}{12}(4+\phi_0)e^{-4B-C} + \ft1{15}e^{-2B+C}
- \ft1{10}e^{-C}\right) \Bigr]\psi^{m=11/2} \nn \\
&& \Bigl[\ft{3i}5e^{-4B}\gamma^\mu\A_\mu 
+ \ft{i}{5}e^{C-A}\gamma^{\mu\nu}F_{\mu\nu}
- \ft{i}{10}e^{-A-2B-C}\gamma^{\mu\nu}p_{\mu\nu}
+ e^A\left(\ft{12}5 e^{-2B+C} - \ft{12}5e^{-C}\right) \Bigr]\psi^{m=-9/2}\nn \\
&& + v_ie^{-2B}\Bigl[\left(- \ft25 \gamma^\mu f_\mu^i
+\ft{34i}{15} e^{A-C}f_0^i\right)\psi^{c\,m=11/2}
+\left(\ft35 \gamma^\mu f_\mu^i
-\ft{7i}5e^{A-C}f_0^i\right)\psi^{c\,m=-9/2}\Bigr] \nn \\
&& + \bar v_ie^{-2B}\left(\ft{3}4\gamma^\mu f^i_\mu
+ \ft{7i}4e^{A-C}f_0^i\right)\lambda', \label{eq:11/2eom}\\
0 &=& \Bigl[\gamma^\mu\mathcal D_\mu + \ft{2i}{5}e^{-4B}\gamma^\mu\A_\mu
-\ft{3i}{40}e^{C-A}\gamma^{\mu\nu}F_{\mu\nu} - \ft{3i}{20}e^{-A-2B-C}\gamma^{\mu\nu}p_{\mu\nu} \nn \\
&&+ e^A\left(\ft{1}{4}(4+\phi_0)e^{-4B-C} + \ft{13}5e^{-2B+C}
 +\ft9{20}e^{-C}\right) \Bigr]\psi^{m=-9/2} \nn \\
&& \Bigl[\ft{2i}5e^{-4B}\gamma^\mu\A_\mu 
+ \ft{2i}{15}e^{C-A}\gamma^{\mu\nu}F_{\mu\nu}
- \ft{i}{15}e^{-A-2B-C}\gamma^{\mu\nu}p_{\mu\nu}
+ e^A\left(\ft{8}5 e^{-2B+C} - \ft{8}5e^{-C}\right) \Bigr]\psi^{m=11/2}\nn \\
&& + v_ie^{-2B}\Bigl[\left(\ft25 \gamma^\mu f_\mu^i
-\ft{14i}{5} e^{A-C}f_0^i\right)\psi^{c\,m=11/2}
+\left(-\ft35 \gamma^\mu f_\mu^i
-\ft{3i}5e^{A-C}f_0^i\right)\psi^{c\,m=-9/2}\Bigr] \nn \\
&& + \bar v_ie^{-2B}\left(\ft{1}2\gamma^\mu f^i_\mu
- \ft{i}2e^{A-C}f_0^i\right)\lambda'.
\label{eq:9/2eom}
\ea
As in the dilatino case, we have defined the supercovariant derivatives
\begin{eqnarray}
\mathcal D_\mu\psi^{m=11/2}&=&\left[\nabla_\mu+\ft{i}{4\tau_2}\partial_\mu
\tau_1-\ft{3i}2(A_\mu+\ft16e^{-4B}\mathbb A_\mu)\right]\psi^{m=11/2}
-K(\psi^{m=11/2})\hat\psi_\mu,\nonumber\\
\mathcal D_\mu\psi^{m=-9/2}&=&\left[\nabla_\mu+\ft{i}{4\tau_2}\partial_\mu
\tau_1-\ft{3i}2(A_\mu+\ft16e^{-4B}\mathbb A_\mu)\right]\psi^{m=-9/2}
-K(\psi^{m=-9/2})\hat\psi_\mu,
\end{eqnarray}
with $K(\psi^{m=11/2})$ and $K(\psi^{m=-9/2})$ similarly obtained
from the variations (\ref{eq:11/2var}) and (\ref{eq:9/2var}), respectively.

Finally, the gravitino equation takes the form
\begin{equation}
0=\gamma^{\mu\nu\rho}\mathcal D_\nu\hat\psi_\rho
-\ft8{15}\tilde K(\psi^{m=11/2})\gamma^\mu\psi^{m=11/2}
-\ft45\tilde K(\psi^{m=-9/2})\gamma^\mu\psi^{m=-9/2}
-\ft12\tilde K(\lambda')\gamma^\mu\lambda',
\end{equation}
where the supercovariant derivative acting on the gravitino is given
by the right hand side of the gravitino variation (\ref{eq:gvar}), and
where the $\tilde K$ terms are essentially the Dirac conjugates of
$K$.
The above equations have the appropriate structure to be obtained from an
effective $\mathcal N=2$ Lagrangian of the form%
\footnote{Note that some care must be taken when considering the conjugate
spinor terms.  Nevertheless, the various conjugate terms do assemble
themselves properly into a consistent effective fermionic Lagrangian.  This
is one place where a more conventional symplectic-Majorana approach would
allow the manipulations to be more transparent.}
\begin{eqnarray}
e^{-1}\mathcal L&=&\bar{\hat\psi}_\mu\gamma^{\mu\nu\rho}\mathcal D_\nu
\hat\psi_\rho+\ft8{15}\bar\psi^{m=11/2}\gamma^\mu D_\mu\psi^{m=11/2}
+\ft45\bar\psi^{m=-9/2}\gamma^\mu D_\mu\psi^{m=-9/2}
+\ft12\bar\lambda'\gamma^\mu D_\mu\lambda'\nonumber\\
&&\!\!\!\!+\left[\bar{\hat\psi}_\mu\left(-\ft8{15}\tilde K(\psi^{m=11/2})\gamma^\mu\psi^{m=11/2}
-\ft45\tilde K(\psi^{m=-9/2})\gamma^\mu\psi^{m=-9/2}
-\ft12\tilde K(\lambda')\gamma^\mu\lambda'\right)
+\mathrm{h.c.}\right]\nonumber\\
&&\!\!\!\!+\cdots.
\end{eqnarray}
The full fermionic Lagrangian (to quadratic order in the fermions) is given
in Appendix~\ref{sec:redlag}.

Although we have worked only to quadratic order in the fermions, it is clear
from the nature of the invariant spinors $\eta$ and $\eta^c$ that higher
spinor harmonics would not be excited by this subset of states.  Thus, if
desired, the consistent truncation may be extended to the four-fermi terms
as well.  However, we expect this to be quite tedious and not particularly
worth pursuing.


\section{A supersymmetric holographic superconductor}
\label{sec:sc}

In this final section we demonstrate the consistency of a particularly
interesting truncation to the lowest Kaluza-Klein level, namely
the supersymmetric completion of the bosonic truncation 
first demonstrated in \cite{Gubser:2009qm}. As we demonstrate, this
is a fully consistent truncation, so long as we keep all fields in
the graviton and
LH+RH chiral multiplets. However, it is a nontrivial truncation, in
that it is not consistent to naively set the other fields in the above
reduction to zero. Instead, the ``backreaction'' on the truncated fields
must be taken into account, effectively setting these modes equal to
something depending on the dynamical fields. Due to this backreaction
on the non dynamical fields, the resulting Lagrangian is nonlinear and so
describes a non-trivial coupling of $\mathcal N = 2$ supergravity
with a single hypermultiplet.

In the bosonic sector the truncation
amounts to keeping only $\{\tau, b^{m^2=-3}\}$ and the graviton
and graviphoton dynamical. In what follows, we will denote
$b^{m^2=-3}$ simply as $b$ so that $(b_0^1,b_0^2) = (b,\tau b)$.
This requires the following constraints on the other terms in the reduction
\cite{Cassani:2010uw,Liu:2010sa}
\begin{equation}
b^{m^2=21}=0,\qquad e^{4B}=e^{-4C} = 1-4\tau_2 |b|^2,\qquad
\mathbb A_1 = -4i\tau_2(bD\bar b - \bar bDb)
+ 4 |b|^2 d \tau_1,
\end{equation}
and
\begin{equation}
\phi_0 = -24\tau_2|b|^2,\qquad p_2 = -d A_1.
\end{equation}

For the fermions, by analyzing the supersymmetry transformations of the
spin-$\fft12$ fields in this truncation, it is evident that if we set 
\beq
\psi = -\psi_9 =-\fft{i}2b\tau_2^{1/2}e^{-2B}\lambda',
\eq 
the resulting system will be consistent with the supersymmetry
transformations. It turns out that under this identification the
fermion equations of motion also degenerate into a single
expression, resulting in a theory containing only $\lambda'$ and
$\hp_\mu$ in the fermionic sector.

Moving directly to the Lagrangian, we write this as a sum of bosonic and fermionic
contributions $\mathcal L = \mathcal L_b + \mathcal L_f$, where
\begin{eqnarray}
\mathcal L_b&=& R*1+ \fft{6(2-3\chi)}{(1-\chi)^2}*1
-\fft{d\chi\wedge*d\chi}{2(1-\chi)^2}  - \fft{(1+\chi)d\tau\wedge*d\bar\tau}{2(1-\chi)\tau_2^2}
 -\fft32F_2\wedge*F_2-\fft{\mathbb A_1\wedge*\mathbb A_1}{2(1-\chi)^2} \nonumber\\
&& - \fft{8\tau_2 Db\wedge*D\bar b}{1-\chi} - \fft{2i}{1-\chi}(\bar b Db\wedge*d\bar\tau - b D\bar b\wedge*d\tau) -A_1\wedge F_2 \wedge F_2,
\end{eqnarray}
and
\bea
e^{-1}\mathcal L_f &=& \bar \hp_\alpha\gamma^{\alpha\beta\sigma}
D_\beta \hp_\sigma +
\ft{3i}8\bar\hp_\alpha\left(\gamma^{\alpha\beta\rho\sigma} +
2g^{\alpha\beta}g^{\rho\sigma}\right)F_{\beta\rho}\hp_\sigma  +
\ft12\bar\tl \gamma^\alpha D_\alpha \tl + 
\ft{3i}{16}\bar\tl\gamma^{\mu\nu}F_{\mu\nu}\tl \nn \\
&& + \ft{1}{2}e^{-4B}\left(3\tau_2(b\bar{
    D_\mu b} - \bar b D_\mu b)\bar\tl\gamma^\mu\tl
  +\ft32(1+8\tau_2|b|^2)\bar\tl\tl\right) \nn\\
&&  +
e^{-4B}\left(-\ft32\bar\hp_\alpha\gamma^{\alpha\sigma}\hp_\sigma
+ \tau_2(\bar bD_\beta b -
b\bar{D_\beta b})\bar\hp_\alpha\gamma^{\alpha\beta\sigma}\hp_\sigma
\right)
\nn \\
&& +  \tau_2^{1/2}e^{-4B}\left(D_\mu b
  \bar\hp_\alpha\gamma^\mu\gamma^\alpha\tl +
  3b\bar\hp_\alpha\gamma^\alpha\tl + h.c.\right) \nn \\
&& +\fft{e^{-2B}}{\tau_2^{1/2}}\left( - b\bar\hp_\alpha\gamma^{\alpha\beta\sigma}
  \partial_\beta\tau\hp_\sigma^c + 
\tau_2^{1/2}\bar\hp_\alpha\gamma^\mu \partial_\mu\tau\gamma^\alpha\tl^c + h.c.\right),
\ea
where we have defined $\tl \equiv e^{-2B}\lambda'$, $\chi =
\tau_2|b|^2$ and we have redefined the gauge covariant derivative acting
on $b$ as
$D_\mu b = \bigl(\partial_\mu  - 3iA_\mu -
\ft{i}{2\tau_2}\partial_\mu\tau_1\bigr)b,$ and similarly for $\tl$ and
$\hp_\alpha$.

This truncation is of interest for many of the condensed
matter applications of the AdS/CFT correspondence involving the coupling of
a charged
scalar and fermion. In particular the original motivation for the
bosonic truncation was in describing a superconducting phase
transition using holographic methods within a controlled system, {\it i.e,}~one
which is derived directly from a UV complete theory. This
truncation hence completes the story by demonstrating the embedding
into a fully supersymmetric theory. It would be interesting to
consider the dynamics of this theory, and whether there is a
supersymmetric superconducting phase transition. Note however that
this analysis would be complicated by the presence of the
gravitino. After all, it is not consistent to simply set the gravitino field
defined here to be zero.  Since the gravitino couples to the supercurrent,
this suggests that the holographic superconductor model of
\cite{Gubser:2009qm} in fact has an underlying (although spontaneously
broken) supersymmetry.

While the truncation first presented in \cite{Gubser:2009qm} did not include
the axi-dilaton, as in the bosonic case, it is consistent to fix
$\tau$ as well.  This simplifies the Lagrangian to be
\begin{eqnarray}
e^{-1}\mathcal L &=& R   - \ft34 F_{\mu\nu}F^{\mu\nu} - e^{-1}A_1\wedge F_2\wedge F_2
\nn \\
&& + 12\fft{(1-6f^2)}{(1-4f^2)^2} - 8\fft{\partial_\mu f\partial^\mu f}{(1-4f^2)^2} - 8f^2\fft{(\partial_\mu \theta -3A_\mu)(\partial^\mu \theta-3A^\mu)}{(1-4f^2)^2} \nn \\
&& + \bar \hp_\alpha\gamma^{\alpha\beta\sigma} D_\beta \hp_\sigma  +
\ft12\bar\tl \gamma^\alpha  D_\alpha \tl +
\ft{3i}8\bar\hp_\alpha\left(\gamma^{\alpha\beta\rho\sigma} + 
2g^{\alpha\beta}g^{\rho\sigma}\right)F_{\beta\rho}\hp_\sigma  +
\ft{3i}{16}\bar\tl\gamma^{\mu\nu}F_{\mu\nu}\tl \nn \\
&& + \fft{1}{1-4f^2}\left(\ft34(1+8f^2)\bar\tl\tl -
\ft32\bar\hp_\alpha\gamma^{\alpha\sigma}\hp_\sigma
-if^2(\partial_\mu\theta-3A_\mu)\left(3\bar\tl\gamma^\mu\tl +
2\bar\hp_\alpha\gamma^{\alpha\beta\sigma}\hp_\sigma\right)\right)  \nn\\
&& + \left(\fft{e^{i\theta}}{1-4f^2}\left(\left(\partial_\mu f+if(\partial_\mu\theta-3 A_\mu)\right)\bar\hp_\alpha\gamma^\mu\gamma^\alpha\tl +
  3f\bar\hp_\alpha\gamma^\alpha\tl\right) + h.c. \right),
\ea
where we have defined $b = \sqrt{g_s}fe^{i\theta}$ and $\tau= ig_s^{-1}$.

Finally, it is worth noting that although this theory involves a charged scalar
coupled to the fermion $\tilde\lambda$, it lacks the Majorana coupling
$\phi\lambda\lambda$ that has been of recent interest in studies involving
gapped fermions in the bosonic condensate
\cite{Chen:2009pt,Faulkner:2009am,Gubser:2009dt}.  While this coupling is
allowed by charge conservation, the explicit reduction shows that it is
not present.  More generally, examination of Table~\ref{tbl:match}
demonstrates that the $b^{m^2=21}$ scalar in the massive vector multiplet
may have such a coupling, and in fact the equations of motion
(\ref{eq:11/2eom}) and (\ref{eq:9/2eom}) show that it is exists for both
$\psi^{m=11/2}$ and $\psi^{m=-9/2}$.  It would be curious to see if this
$b^{m^2=21}$ scalar may play a role in novel models of holographic
superconductors.

\begin{acknowledgments}

We wish to thank I. Bah, A. Faraggi, N. Halmagyi and D. Vaman for useful
discussions.
This work was supported in part by the US Department of Energy under grant
DE-FG02-95ER40899.

\end{acknowledgments}

\appendix

\section{Dirac Matrix Conventions} \label{app:dirac}

We work with a mostly plus metric signature, and take the conventional
Clifford algebra $\{\Gamma^A,\Gamma^B\}=2\eta^{AB}$.  Note, in particular,
that $\Gamma^0$ is anti-hermitian, so that $(\Gamma^0)^\dagger=-\Gamma^0$
and $(\Gamma^i)^\dagger=\Gamma^i$.  The ten-dimensional Chirality matrix
is given by
\begin{equation}
\Gamma^{11}\equiv\fft1{10!}\epsilon_{A_1\cdots A_{10}}\Gamma^{A_1}\cdots
\Gamma^{A_{10}}=\Gamma^0\cdot\cdot\cdot\Gamma^9,
\end{equation}
and squares to the identity.

Corresponding to the metric reduction (\ref{eq:metans}), we decompose the
ten-dimensional Dirac matrices according to
\bea
\Gamma^{\alpha}& \equiv & \gamma^\alpha \otimes 1_4 \otimes \sigma_1,\nonumber\\
\Gamma^{a}& \equiv & 1_4\otimes \tau^a \otimes \sigma_2,\nonumber\\
\Gamma^9 &\equiv & 1_4\otimes \tau^9 \otimes \sigma_2,
\ea
where $\gamma^\alpha$ are Dirac matrices in the five-dimensional spacetime with $\gamma^4 \equiv i\gamma^0\gamma^1\gamma^2\gamma^3$ and $\tau^a$ are Dirac matrices in the five-dimensional internal space with $\tau^9 \equiv \tau^5\tau^6\tau^7\tau^8$.  The
Chirality matrix $\Gamma^{11}$ is then given by
\begin{equation}
\Gamma^{11} = \Gamma^0\cdot\cdot\cdot\Gamma^9 = 1_4\otimes1_4\otimes\sigma_3.
\end{equation}

We furthermore take the following conventions for the $A$, $C$ and $D$
intertwiners which map between different representations of the Dirac matrices
\beq
A_{10}\Gamma_M A_{10}^{-1}=\Gamma_M^\dagger, \qquad C_{10}^{-1}\Gamma_M C_{10}=-\Gamma_M^T, \qquad D_{10}^{-1}\Gamma_M D_{10}=-\Gamma_M^*.
\eq
Here $C_{10}$ denotes the charge conjugation matrix. These may be decomposed
as
\beq
A_{10} = A_{4,1}\otimes A_5\otimes\sigma_1, \qquad C_{10} = C_{4,1}\otimes C_5\otimes \sigma_2, \qquad D_{10} = iD_{4,1}\otimes D_5\otimes \sigma_3,
\eq
where the five-dimensional intertwiners are defined as
\bea
 A_{4,1}\gamma_\mu A_{4,1}^{-1} = -\gamma_\mu^\dagger, &\qquad  C_{4,1}^{-1}\gamma_\mu C_{4,1} = \gamma_\mu^T, & \qquad D_{4,1}^{-1}\gamma_\mu D_{4,1} = -\gamma_\mu^* \nn \\
 A_5\tau_a A_5^{-1} = \tau_a^\dagger, & \qquad C_5^{-1}\tau_a C_5 = \tau_a^T, & \qquad D_5^{-1}\tau_a D_5 = \tau_a^*.
\ea
It turns out the following is a consistent decomposition:
\beq
A_{10} = \Gamma_0 = \gamma_0\otimes1\otimes\sigma_1, \qquad C_{10} = C_{4,1}\otimes C_5\otimes \sigma_2, \qquad D_{10} = i\gamma_0C_{4,1}\otimes C_5\otimes \sigma_3.
\eq
The five dimensional charge conjugation matrices on both spacetime and the
internal manifold satisfy
\beq
C_5 = -C_5^T = C_5^* = -C_5^{-1}.
\eq

Finally, we define the charge conjugate of a spinor in any dimension to
be $\psi^c = C A^T \psi^*$, which is equivalent to
$\psi^c = -\Gamma_0 C_{10}\psi^*$.  Therefore, letting $\chi$ and $\eta$
be spinors on $M$ and $SE_5$, respectively, the charge conjugates are given
by $\chi^c = -\gamma_0C_{4,1}\chi^*$ and $\eta^c = C_5 \eta^*$.

\section{The Reduced Lagrangian}
\label{sec:redlag}

The bosonic Lagrangian with the massive gravitino multiplet removed was
presented in \cite{Liu:2010sa}, and takes the form
\begin{eqnarray}
\mathcal L_b&=&R*1+(24e^{2A-2B}-4e^{5A+3C}-\ft12e^{8A}(4+\phi_0)^2)*1
-\ft{28}3dB\wedge*dB-\ft83dB\wedge*dC\nonumber\\
&&-\ft43dC\wedge*dC-\ft1{2\tau_2^2}d\tau\wedge*d\bar\tau
-\ft12e^{2C-2A}F_2\wedge*F_2-e^{A-C}(F_2+\ft14\mathbb F'_2)\wedge*
(F_2+\ft14\mathbb F'_2)\nonumber\\
&&-\ft12e^{-8B}[\mathbb A'_1-\ft{2i}3\epsilon_{ij}(f_0^i\bar f_1^j
-\bar f_0^if_1^j)]\wedge*[\mathbb A'_1
-\ft{2i}3\epsilon_{ij}(f_0^i\bar f_1^j-\bar f_0^if_1^j)]\nonumber\\
&&-2\mathcal M_{ij}
[e^{5A-C}(f_0^i\bar f_0^j+\bar f_0^if_0^j)*1
+e^{-4B}(f_1^i\wedge*\bar f_1^j+\bar f_1^i\wedge*f_1^j)]\nonumber\\
&&-A_1\wedge (F_2+\ft14\mathbb F'_2)\wedge (F_2+\ft14\mathbb F'_2),
\label{eq:non=1}
\end{eqnarray}
where $\mathbb A_1'=\mathbb A_1+\ft{2i}3\epsilon_{ij}(f_0^i\bar f_1^j
-\bar f_0^if_1^j)$, and where $\mathbb F_2'=d\mathbb A_1'$.

The corresponding fermionic Lagrangian may be obtained from the equations
of motion presented in Section~\ref{sec:iibeom}.  At quadratic order in
the fermions, we have
\begin{eqnarray}
e^{-1}\mathcal L_f&=& \bar\hp_\mu\gamma^{\mu\nu\rho}\mathcal D_\nu\hp_\rho \nn \\
&&+ \Bigl[-\ft8{15}\bar\psi^{\ma}\gamma^\mu K(\psi^{\ma})\hp_\mu -
\ft45 \bar\psi^{\mb}\gamma^\mu K(\psi^{\mb})\hp_\mu\nonumber\\
&&\kern8em -
\ft12\bar\lambda'\gamma^\mu K(\lambda')\hp_\mu + h.c. \Bigr] \nn\\
&& + \ft8{15} \bar\psi^{\ma} \Bigl[\gamma^\mu D_\mu +
  \ft{3i}{5}e^{-4B}\gamma^\mu\A_\mu 
-\ft{i}{120}e^{C-A}\gamma^{\mu\nu}F_{\mu\nu} -
\ft{11i}{60}e^{-A-2B-C}\gamma^{\mu\nu}p_{\mu\nu} \nn \\ 
&&\kern8em + e^A\left(-\ft{17}{12}(4+\phi_0)e^{-4B-C} + \ft1{15}e^{-2B+C}
- \ft1{10}e^{-C}\right) \Bigr]\psi^{m=11/2} \nn \\
&& +\ft45 \psi^{\mb}\Bigl[\gamma^\mu D_\mu + \ft{2i}{5}e^{-4B}\gamma^\mu\A_\mu
-\ft{3i}{40}e^{C-A}\gamma^{\mu\nu}F_{\mu\nu} -
\ft{3i}{20}e^{-A-2B-C}\gamma^{\mu\nu}p_{\mu\nu} \nn \\ 
&&\kern8em + e^A\left(\ft{1}{4}(4+\phi_0)e^{-4B-C} + \ft{13}5e^{-2B+C}
 +\ft9{20}e^{-C}\right) \Bigr]\psi^{m=-9/2} \nn \\
&& + \ft12\bar\lambda' \Bigl[\gamma^\mu D_\mu+\ft{i}8\gamma^{\mu\nu}
\left(e^{C-A}F_{\mu\nu}-2e^{-A-2B-C}p_{\mu\nu}\right) \nn \\
&&\kern8em - \ft14(4+\phi_0)e^{A-4B-C} +
e^{A-2B+C} + \ft32e^{A-C}\Bigr]\lambda' \nn \\
&& + \ft8{15}\Bigl[ \bar\psi^{\ma}\Bigl(\ft{3i}5e^{-4B}\gamma^\mu\A_\mu 
+ \ft{i}{5}e^{C-A}\gamma^{\mu\nu}F_{\mu\nu}
- \ft{i}{10}e^{-A-2B-C}\gamma^{\mu\nu}p_{\mu\nu}\nn \\
&&\kern8em + e^A\left(\ft{12}5 e^{-2B+C} - \ft{12}5e^{-C}\right)
\Bigr)\psi^{m=-9/2} + h.c.\Bigr]\nn \\
&& + \ft8{15}\Bigl[ v_ie^{-2B}\bar\psi^{\ma} \left(- \ft25 \gamma^\mu f_\mu^i
+ \ft{34i}{15} e^{A-C}f_0^i\right)\psi^{c\,m=11/2} + h.c. \Bigr] \nn\\
&& + \ft8{15} \Bigl[ v_ie^{-2B} \bar\psi^{\ma}\left( \ft35 \gamma^\mu f_\mu^i
-\ft{7i}5e^{A-C}f_0^i\right)\psi^{c\,m=-9/2} + h.c. \Bigr] \nn \\
&&+\ft45\Bigl[v_ie^{-2B}\bar\psi^{\mb}\left(-\ft35 \gamma^\mu f_\mu^i
-\ft{3i}5e^{A-C}f_0^i\right)\psi^{c\,m=-9/2} + h.c.\Bigr] \nn \\
&& + \ft8{15}\Bigl[ \bar v_ie^{-2B} \bar\psi^{\ma}\left(\ft{3}4\gamma^\mu f^i_\mu
+ \ft{7i}4e^{A-C}f_0^i\right)\lambda' + h.c. \Bigr] \nn\\
&& +\ft45\Bigl[ \bar v_ie^{-2B}\bar\psi^{\mb}\left(\ft{1}2\gamma^\mu
   f^i_\mu - \ft{i}2e^{A-C}f_0^i\right)\lambda' + h.c.\Bigr],
\end{eqnarray}
and the full Lagrangian up to quadratic order in the fermions is given
by
\beq
\mathcal L = \mathcal L_b +\mathcal L_f.
\eq



\begin{thebibliography}{99}

\bibitem{Bremer:1998zp}
M.~S.~Bremer, M.~J.~Duff, H.~Lu, C.~N.~Pope and K.~S.~Stelle,
{\sl Instanton cosmology and domain walls from M-theory and string theory},
Nucl.\ Phys.\  B {\bf 543}, 321 (1999) [arXiv:hep-th/9807051].

\bibitem{Liu:2000gk}
J.~T.~Liu and H.~Sati,
{\sl Breathing mode compactifications and supersymmetry of the brane-world},
Nucl.\ Phys.\  B {\bf 605}, 116 (2001) [arXiv:hep-th/0009184].

\bibitem{Gauntlett:2009zw}
J.~P.~Gauntlett, S.~Kim, O.~Varela and D.~Waldram,
{\sl Consistent supersymmetric Kaluza--Klein truncations with massive modes},
JHEP {\bf 0904}, 102 (2009) [arXiv:0901.0676 [hep-th]].

\bibitem{Cassani:2010uw}
D.~Cassani, G.~Dall'Agata and A.~F.~Faedo,
{\sl Type IIB supergravity on squashed Sasaki-Einstein manifolds},
JHEP {\bf 1005}, 094 (2010) [arXiv:1003.4283 [hep-th]].

\bibitem{Liu:2010sa}
J.~T.~Liu, P.~Szepietowski and Z.~Zhao,
{\sl Consistent massive truncations of IIB supergravity on Sasaki-Einstein
manifolds},
Phys.\ Rev.\  D {\bf 81}, 124028 (2010) [arXiv:1003.5374 [hep-th]].

\bibitem{Gauntlett:2010vu}
J.~P.~Gauntlett and O.~Varela,
{\sl Universal Kaluza-Klein reductions of type IIB to $\mathcal N=4$
supergravity in five dimensions},
JHEP {\bf 1006}, 081 (2010) [arXiv:1003.5642 [hep-th]].

\bibitem{Skenderis:2010vz}
K.~Skenderis, M.~Taylor and D.~Tsimpis,
{\sl A consistent truncation of IIB supergravity on manifolds admitting a
Sasaki-Einstein structure},
JHEP {\bf 1006}, 025 (2010) [arXiv:1003.5657 [hep-th]].

\bibitem{Cassani:2010na}
D.~Cassani and A.~F.~Faedo,
{\sl A supersymmetric consistent truncation for conifold solutions},
arXiv:1008.0883 [hep-th].

\bibitem{Bena:2010pr}
I.~Bena, G.~Giecold, M.~Grana, N.~Halmagyi and F.~Orsi,
{\sl Supersymmetric Consistent Truncations of IIB on $T^{1,1}$},
arXiv:1008.0983 [hep-th].

\bibitem{Gauntlett:2009dn}
J.~P.~Gauntlett, J.~Sonner and T.~Wiseman,
{\sl Holographic superconductivity in M-Theory},
Phys.\ Rev.\ Lett.\  {\bf 103}, 151601 (2009) [arXiv:0907.3796 [hep-th]].

\bibitem{Gauntlett:2009bh}
J.~P.~Gauntlett, J.~Sonner and T.~Wiseman,
{\sl Quantum Criticality and Holographic Superconductors in M-theory},
JHEP {\bf 1002}, 060 (2010) [arXiv:0912.0512 [hep-th]].

\bibitem{Gubser:2009qm}
S.~S.~Gubser, C.~P.~Herzog, S.~S.~Pufu and T.~Tesileanu,
{\sl Superconductors from Superstrings},
Phys.\ Rev.\ Lett.\  {\bf 103}, 141601 (2009) [arXiv:0907.3510 [hep-th]].

\bibitem{Bah:2010yt}
I.~Bah, A.~Faraggi, J.~I.~Jottar, R.~G.~Leigh and L.~A.~P.~Zayas,
{\sl Fermions and $D=11$ Supergravity On Squashed Sasaki-Einstein Manifolds},
arXiv:1008.1423 [hep-th].

\bibitem{Bah:2010cu}
I.~Bah, A.~Faraggi, J.~I.~Jottar and R.~G.~Leigh,
{\sl Fermions and Type IIB Supergravity On Squashed Sasaki-Einstein Manifolds},
arXiv:1009.1615 [hep-th].

\bibitem{Schwarz:1983qr}
J.~H.~Schwarz,
{\sl Covariant Field Equations Of Chiral $N=2$ $D=10$ Supergravity},
Nucl.\ Phys.\  B {\bf 226}, 269 (1983).


\bibitem{Gibbons:2002th}
G.~W.~Gibbons, S.~A.~Hartnoll and C.~N.~Pope,
{\sl Bohm and Einstein-Sasaki metrics, black holes and cosmological event
horizons},
Phys.\ Rev.\  D {\bf 67}, 084024 (2003)
[arXiv:hep-th/0208031].

\bibitem{Gunaydin:1984fk}
M.~Gunaydin and N.~Marcus,
{\sl The Spectrum Of The $S^5$ Compactification Of The Chiral $N=2$, $D=10$
Supergravity And The Unitary Supermultiplets Of U(2,2/4)},
Class.\ Quant.\ Grav.\  {\bf 2}, L11 (1985).

\bibitem{Kim:1985ez}
H.~J.~Kim, L.~J.~Romans and P.~van Nieuwenhuizen,
{\sl The Mass Spectrum Of Chiral $N=2$ $D=10$ Supergravity On $S^5$},
Phys.\ Rev.\  D {\bf 32}, 389 (1985).

\bibitem{Chen:2009pt}
J.~W.~Chen, Y.~J.~Kao and W.~Y.~Wen,
{\sl Peak-Dip-Hump from Holographic Superconductivity},
Phys.\ Rev.\  D {\bf 82}, 026007 (2010) [arXiv:0911.2821 [hep-th]].

\bibitem{Faulkner:2009am}
T.~Faulkner, G.~T.~Horowitz, J.~McGreevy, M.~M.~Roberts and D.~Vegh,
{\sl Photoemission 'experiments' on holographic superconductors},
JHEP {\bf 1003}, 121 (2010) [arXiv:0911.3402 [hep-th]].

\bibitem{Gubser:2009dt}
S.~S.~Gubser, F.~D.~Rocha and P.~Talavera,
{\sl Normalizable fermion modes in a holographic superconductor},
arXiv:0911.3632 [hep-th].



\end{thebibliography}
\end{document}